\documentclass[aps,pra,twocolumn,superscriptaddress,amsmath,amssymb,showpacs,notitlepage]{revtex4-1}
\pdfoutput=1
\usepackage{algorithm}
\usepackage{color}
\usepackage{algorithmic}
\usepackage{amsmath}
\usepackage{braket}
\usepackage{multirow}
\usepackage{amssymb}
\usepackage{amsthm}
\usepackage{bm}
\usepackage{thmtools}

\usepackage{array}
\usepackage{url}
\usepackage{hyperref}
\hypersetup{
    colorlinks=true,
    linkcolor=blue,
    filecolor=magenta,      
    urlcolor=cyan,
    citecolor=red
}
\usepackage{graphicx}
\usepackage{bibentry}
\usepackage{booktabs}
\usepackage{dcolumn}
\usepackage{bm}
\usepackage{blkarray}
\usepackage{xparse}
\usepackage{mathtools}
\usepackage{microtype}
\usepackage{enumerate}
\usepackage{soul}

\providecommand{\U}[1]{\protect\rule{.1in}{.1in}}
\newtheorem*{theorem*}{Theorem}

\begin{document}

\title{Bayesian minimum mean square error for transmissivity sensing}

\author{Boyu Zhou}
\affiliation{Department of Physics, The University of Arizona
1118 E. Fourth Street, Tucson, AZ 85721, USA}
\author{Boulat A. Bash}
\affiliation{Department of Electrical and Computer Engineering, The University of Arizona, Tucson AZ, 85721, USA}
\affiliation{Wyant College of Optical Sciences, The University of Arizona, 1630 E. University Blvd., Tucson, AZ 85721, USA}
\author{Saikat Guha}
\affiliation{Wyant College of Optical Sciences, The University of Arizona, 1630 E. University Blvd., Tucson, AZ 85721, USA}
\affiliation{Department of Electrical and Computer Engineering, The University of Arizona, Tucson AZ, 85721, USA}
\author{Christos N. Gagatsos}
\affiliation{Department of Electrical and Computer Engineering, The University of Arizona, Tucson AZ, 85721, USA}
\affiliation{Wyant College of Optical Sciences, The University of Arizona, 1630 E. University Blvd., Tucson, AZ 85721, USA}

\begin{abstract}
We address the problem of estimating the transmissivity of the pure-loss channel from the Bayesian point of view, i.e., we consider that some prior probability distribution function (PDF) on the unknown variable is available and we employ methods to compute the Bayesian minimum mean square error (MMSE). Specifically, we consider two prior PDFs: the two-point and the beta distributions. By fixing the input mean photon number to an integer, for the two-point PDF we prove analytically that the optimal state is the Fock state and the optimal measurement is photon-counting, while for the beta PDF our numerical investigation provides evidence on the optimality of the Fock state and photon-counting. Moreover, we investigate the situation where the input mean photon number is any (non-negative) real number. For said case, we conjecture the form of the optimal input states and we study the performance of photon-counting, which is a sub-optimal yet practical measurement. Our methods can be applied for any prior PDF. We emphasize that we compute the MMSE instead of Bayesian lower bounds on the mean square error based on the Fisherian approach.
\end{abstract}
\maketitle

\section{Introduction}
Estimation of the transmissivity $\tau\in [0,1]$ (or equivalently of loss) is of paramount importance in quantum 
communications as it determines the fundamental limit of 
any point-to-point quantum protocol \cite{Pirandola2017}. In the quantum sensing setting, estimation of loss of power is connected to the reflectance of a target, and it affects the sensing system's ability to attain Heisenberg scaling in many scenarios \cite{Demkowicz2012,Escher2011}. Moreover, transmissivity estimation plays a crucial role in quantum network diagnostics: Given a network where information-carrying light is transmitted through, one wishes to estimate what the unavoidable losses are due to imperfections, for experimental, manufacturing, or other practical reasons. Lastly, estimation of transmissivity and the questions that come with it (e.g. optimal states and measurements for said task) is interesting in its own right from the theory point of view.

To our knowledge, the problem at hand has been addressed through the Fisherian approach \cite{Nair2018,Adesso2009,Monras2007,Gong2022,Jonsson2022}, i.e., considering the quantum and classical Fisher information (denoted as $F(\tau)$ and $C(\tau)$) whose inverses serve as lower bounds on the mean square error (MSE), with $F^{-1}(\tau)$ being the fundamental limit. These bounds are the well-known Cram\'er-Rao bounds. In this work, we do not consider the unknown parameter as having a fixed value, but we consider it as being a random variable that follows a prior probability distribution (PDF) $P(\tau)$, which gives rise to Bayesian quantum estimation \cite{Personick1971,Rubio2020}. The use of a prior PDF can be justified as a result of prior measurements of the system \cite{Lee2022} or of being based upon natural assumptions, e.g., the manufacturer can provide information, in the form of a PDF, on the loss of an optical fiber by performing diagnostics on random samples, but without specifying the loss rate of the equipment at hand. 

The pure-loss channel is a successful model for lossy optical elements and is defined as a beam splitter with transmissivity $\tau$, whose lower input mode is set to vacuum and its lower output mode is traced out. The upper input and output modes are referred to as input and output of the channel respectively. For said channel, and within the Fisherian approach, one of the results in \cite{Adesso2009} was that for fixed input mean photon number, the Fock state is the optimal probe state to sense $\tau$ when the input mean photon number is an integer, while for that case photon-counting was proven to be the optimal measurement. Results for fixed mean photon number to non-integers were also derived. 

If a prior PDF is available, one can construct Bayesian versions of the Fisherian Cram\'er-Rao bounds: Let us consider a single unknown variable $x$ that follows a (prior) PDF $P(x)$. Also, let us consider the optimal measurement in the sense that the QFI is equal to the CFI, i.e., $F(x)=C(x)$, where we note that such a measurement always exists in the single-parameter setting. Then, the Bayesian version of the Cram\'er-Rao bound is \cite{VanTrees2013} ,
\begin{eqnarray}
 \label{eq:BayesianCRB} \delta_B \geq J_E^{-1} \geq J_B^{-1}, 
\end{eqnarray}
where,
\begin{eqnarray}
  \label{eq:deltaB}  \delta_B = \int dx P(x) \text{tr}\left[\hat{\rho}(x)(\hat{H}-x\hat{I})^2\right]
\end{eqnarray}
is the Bayesian mean square error for a measurement given by the eigenvectors of the Hermitian operator $\hat{H}$ on the final state $\hat{\rho}(x)$,
\begin{eqnarray}
 \label{eq:JEinv}  J_E^{-1} &=& \int dx P(x) F^{-1}(x),\\
 \label{eq:JB}  J_B &=& J_D + J_P,
\end{eqnarray}
where $J_D$ is the Fisher information related to data,
\begin{eqnarray}
 \label{eq:JD}   J_D = \int dx P(x) F(x)
\end{eqnarray}
and $J_P$ is the Fisher information related to the prior PDF,
\begin{eqnarray}
  \label{eq:JP}  J_P = \int dx P(x) \left(\frac{\partial \ln P(x)}{\partial x}\right)^2.
\end{eqnarray}
In inequality \eqref{eq:BayesianCRB}, $J_B$ is not always attainable and $J_E^{-1}$ is attainable in the limit where the collected data goes to infinity \cite{VanTrees2013}. Therefore, said bounds are not guaranteed to fully capture the impact of prior information. For any given Bayesian sensing task, even if examined using the Fisherian approach, a rigorous analysis must be re-done in a genuinely Bayesian fashion \cite{Personick1971,Rubio2020}. However, some physical intuition can be drawn from results obtained in the Fisherian approach.

In this paper we set $x=\tau$ and we consider the formalism developed in \cite{Personick1971}: We work with the minimum mean square error (MMSE) where the minimization is performed over all possible protective measurements, i.e., all possible Hermitian operators $\hat{H}$ of Eq. \eqref{eq:deltaB},
\begin{eqnarray}
\delta \equiv \underset{\hat{H}}{\text{min}}\ \delta_B.
\end{eqnarray}

This paper is organized as follows: In Section \ref{sec:math} we describe the pure-loss channel and review the Bayesian approach to single-variable quantum estimation  \cite{Personick1971}. In Section \ref{sec:lb} we derive a lower bound on the MMSE which will be useful to the technical development of later sections. In Section \ref{sec:int} we prove that for fixed input mean photon number to an integer and for the two-point PDF as prior, the Fock state is optimal and also find that the optimal measurement is photon-counting. In Section \ref{sec:non-int}, we consider the two-point prior PDF and we do a numerical investigation on what states are optimal for non-integer fixed input mean photon and we examine the behavior of photon-counting as a feasible (yet sub-optimal) measurement. In Section \ref{sec:beta}, we repeat the same analysis, this time based only on numerics, for the beta distribution and we find similar results to the previous sections. Finally, in Section \ref{sec:concl}, we discuss further our results, we do a comparison between the bounds of inequality \eqref{eq:BayesianCRB} and the MMSE $\delta$, and we discuss future research directions.

\section{The mathematical tools and the setup}\label{sec:math}
We denote the input of the pure-loss channel as $\hat{\rho}_0$. Then, the output state is $\hat{\rho}(\tau)=\mathcal{L}_\tau[\hat{\rho}_0]$. The time evolution of the input state $\hat{\rho}_0$ is governed by the following master equation and its initial condition,
\begin{eqnarray}
\label{eq:MasterEq}\frac{\partial \hat{\rho}(t)}{\partial t} &=& \frac{\gamma}{2}[2 \hat{a} \hat{\rho}(t) \hat{a}^\dagger-\hat{a}^\dagger\hat{a} \hat{\rho}(t)-\hat{\rho}(t)\hat{a}^\dagger\hat{a}],\\
\label{eq:InitialCond}\hat{\rho}(t=0)&=&\hat{\rho}_0,
\end{eqnarray}
where the $\hat{a}$, $\hat{a}^\dagger$ are respectively the annihilation and creation operators of a single bosonic mode satisfying the commutation relation $[\hat{a},\hat{a}^\dagger]=1$ and through which we define the number operator $\hat{n}=\hat{a}^\dagger \hat{a}$. The loss parameter $\gamma$ is related to the transmissivity $\tau$ of the pure-loss channel as follows,
\begin{eqnarray}
\label{eq:Tau}    \tau = \exp(-\gamma t).
\end{eqnarray}
Under the parameterization of Eq. \eqref{eq:Tau}, the solution to Eq. \eqref{eq:MasterEq} under the initial condition of Eq. \eqref{eq:InitialCond} is,
\begin{eqnarray}
\label{eq:Kraus}    \hat{\rho}(\tau)= \sum_{l=0}^{\infty} \frac{(1-\tau)^l}{l!}\sqrt{\tau}^{\hat{n}} \hat{a}^l \hat{\rho}_0 \hat{a}^{\dagger l} \sqrt{\tau}^{\hat{n}}.
\end{eqnarray}

The problem set up we consider in this work is as follows: We consider pure input states $\hat{\rho}_0=|\Psi_{\bar{n}}\rangle\langle \Psi_{\bar{n}}|$ with fixed mean-photon number $\langle\Psi_{\bar{n}}|\hat{n}|\Psi_{\bar{n}}\rangle=\bar{n}$. We consider as unknown parameter the transmissivity $\tau$ and on top of this we assume that $\tau$ is a random variable following a prior PDF, $P(\tau)$. Then, the task is to find the optimal input state that minimizes the minimum mean square error (MMSE). Under this Bayesian approach, Personick derived the formulas that give the MMSE $\delta$ and which projective measurement attains it \cite{Personick1971}. For our purposes, said formulas are written as,
\begin{eqnarray}
\label{eq:delta1}\delta = \text{tr}\hat{\Gamma}_2 - \text{tr}(\hat{B} \hat{\Gamma}_1).
\end{eqnarray}
where,
\begin{eqnarray}
  \label{eq:Gammak}  \hat{\Gamma}_k = \int_0^1 d\tau P(\tau) \tau^k \hat{\rho}(\tau),\ k=0,1,2.
\end{eqnarray}
and
\begin{eqnarray}
\label{eq:B}\hat{B} = 2 \int_0^\infty dze^{-z \hat{\Gamma}_0} \hat{\Gamma}_1 e^{-z \hat{\Gamma}_0},
\end{eqnarray}
where the MMSE $\delta$ is always attainable and the optimal projective measurement is given by the eigenvectors of $\hat{B}$.

In this work we specifically consider the two-point prior PDF,
\begin{eqnarray}
\label{eq:TwoPointPrior}P(\tau)=q \delta(\tau-\tau_0)+(1-q) \delta(\tau-\tau_1),
\end{eqnarray}
where $0\leq q \leq 1$, i.e., the transmissivity can take two possible values $\tau_0$ and $\tau_1$ with probabilities $q$ and $1-q$ respectively. The two-point prior PDF allows for two possible transmissivities, occurring with a probabilities $q$ and $1-q$. This problem is akin to target detection and quantum reading but viewed from the angle of the MMSE instead of the minimum probability of error \cite{Pirandola2011,Nair2011}.

We also discuss briefly the beta distribution
\begin{eqnarray}
\label{eq:betaPrior}P(\tau)=\frac{\tau^{\alpha-1}(1-\tau)^{\beta-1}}{\mathcal{B}(\alpha,\beta)},
\end{eqnarray}
where $\mathcal{B}(\alpha,\beta)=\Gamma(\alpha)\Gamma(\beta)/\Gamma(\alpha+\beta)$, $\Gamma(.)$ is the gamma function, and the values of $\alpha>0$, $\beta>0$ cover a great variety of PDF's behavior.

We note here that if the prior PDF is simply a Dirac delta $P(\tau)=\delta(\tau-\tau_0)$ then one can prove that the MMSE is $\delta=0$, which makes perfect sense since such a prior PDF gives full knowledge on the parameter.

Also, we note that in our analysis we assume that the optimal input states are pure since classical mixing involves throwing away information.

\section{Lower bound on the MMSE}\label{sec:lb}
The first term of Eq. \eqref{eq:delta1} depends only on the prior PDF, i.e., since $\text{tr}\hat{\rho}=1$, $\text{tr}\hat{\Gamma}_2=\int_0^1 P(\tau) \tau^2$. Since we fix the prior PDF, when lower-bounding the MMSE we only consider the input state which affects the second term of Eq. \eqref{eq:delta1}. 

Upper bounding the second term of Eq. \eqref{eq:delta1} will yield a lower bound on $\delta$.
We write the second term of Eq. \eqref{eq:delta1} as,
\begin{eqnarray}
\text{tr}(\hat{B} \hat{\Gamma}_1) = 2 \int_0^\infty dz\ \text{tr}\left(e^{-z \hat{\Gamma}_0} \hat{\Gamma}_1 e^{-z \hat{\Gamma}_0} \hat{\Gamma}_1\right).
\end{eqnarray}
For Hermitian operators $\hat{X}$ and $\hat{Y}$, we have the inequality $\text{tr}(\hat{X}\hat{Y}\hat{X}\hat{Y})\leq \text{tr}(\hat{X}^2\hat{Y}^2)$ \cite{Chang1999}. The operators $\hat{\Gamma}_k$ are Hermitian, therefore using said inequality we write,
\begin{eqnarray}
\label{eq:UB1}\text{tr}(\hat{B} \hat{\Gamma}_1) &\leq& 2\int_0^\infty dz\ \text{tr}\left(e^{-2 z \hat{\Gamma}_0} \hat{\Gamma}_1^2\right) \Leftrightarrow\\
\label{eq:UB2}\text{tr}(\hat{B} \hat{\Gamma}_1) &\leq& \text{tr}(\hat{\Gamma}_0^{-1}\hat{\Gamma}_1^2)=\text{tr}(\hat{\Gamma}_1^2 \hat{\Gamma}_0^{-1}),
\end{eqnarray}
where this lower bound is meaningful as long as $\hat{\Gamma}_0$ is invertible. Under said condition, we get a lower bound on the MMSE $\delta$,
\begin{eqnarray}
\label{eq:LB}\delta &\geq& \delta_{\text{LB}},\\
\label{eq:deltaLB} \delta_{\text{LB}} &=& \int_0^1 d\tau P(\tau) \tau^2  - \text{tr}(\hat{\Gamma}_0^{-1}\hat{\Gamma}_1^2).
\end{eqnarray}
In inequality \eqref{eq:LB}, equality is achieved if and only if $[\hat{\Gamma}_0,\hat{\Gamma}_1]=0$.

\section{Integer mean photon number: Optimality of Fock states}\label{sec:int}
In this section we prove that for input mean photon number fixed to an integer, i.e., $\bar{n}=n\in \mathbb{N}$, and for the prior PDF of Eq. \eqref{eq:TwoPointPrior}, the only pure state that attains the lower bound of Eq. \eqref{eq:deltaLB} is the the Fock state $|n\rangle$, i.e., the Fock state is optimal in the sense that attains the lowest MMSE. For this case, we also prove that the optimal measurement is photon-counting or photon number resolving (PNR) detection, i.e., projection on the Fock basis.

Using Eq. \eqref{eq:Gammak}, the condition $[\hat{\Gamma}_0,\hat{\Gamma}_1]=0$ implies,
\begin{eqnarray}
 \label{eq:comm1}   
 \int_0^1 d\tau \int_0^1 d\tau' \tau' P(\tau) P(\tau') [\hat{\rho}(\tau),\hat{\rho}(\tau')]=0.
\end{eqnarray}
For the prior of Eq. \eqref{eq:TwoPointPrior}, Eq. \eqref{eq:comm1} gives,
\begin{eqnarray}
    \label{eq:comm2} q (1-q) (\tau_0-\tau_1) [\hat{\rho}(\tau_0),\hat{\rho}(\tau_1)]=0,
\end{eqnarray}
where the integrals disappeared because of the Dirac deltas, and two terms gave zero as they are proportional to $[\hat{\rho}(\tau_0),\hat{\rho}(\tau_0)]=0$ and $[\hat{\rho}(\tau_1),\hat{\rho}(\tau_1)]=0$.
From Eq. \eqref{eq:comm2}, there are three options: (i) $q=0$ or $q=1$, which means that the prior is simply a single Dirac delta and for that case we get that the MMSE is $\delta=0$ regardless of the input state as full information on the parameter is provided by the prior PDF, (ii) $\tau_0=\tau_1$, in which case the prior PDF is again a single Dirac delta, and (iii) $[\hat{\rho}(\tau_0),\hat{\rho}(\tau_1)]=0$.

Since $\tau_0$ and $\tau_1$ are arbitrary values of the transmissivity, the non-trivial option $[\hat{\rho}(\tau_0),\hat{\rho}(\tau_1)]=0$ means that any two outputs of the pure-loss channel must commute so that the lower bound of Eq. \eqref{eq:deltaLB} is attained.

It is known that if the input of the pure-loss channel is a Fock state, the output will remain Fock diagonal. Therefore, Fock states satisfy the condition $[\hat{\Gamma}_0,\hat{\Gamma}_1]=0$ and as a result they attain the lower bound of Eq. \eqref{eq:deltaLB}.

At the same time, the condition $[\hat{\Gamma}_0,\hat{\Gamma}_1]=0$ must be true for all $\tau_0$ and $\tau_1$, including the choice $\tau_0=0$ and $0<\tau_1\leq 1$. For said choice, from Eq. \eqref{eq:Kraus} we get $\hat{\rho}(\tau_0=0)=|0\rangle \langle 0 |$, which means that $\hat{\rho}(\tau_1)$ must be Fock-diagonal for all $0<\tau_1\leq 1$. In Appendix \ref{app:FockDiagonal} we prove that a Fock-diagonal output of the pure-loss channel can only correspond to a Fock-diagonal input, and since we restrict our search to pure states, the only Fock-diagonal input state is the Fock state.

Furthermore, let a state $|\Psi_n\rangle$ and a Fock state $|n\rangle$, with mean photon number $\langle\Psi_n| \hat{n}|\Psi_n\rangle=\langle n| \hat{n}|n\rangle=n$. Since all states in a Hilbert space can be connected through at least one unitary operator, there is always at least one unitary $\hat{U}_n$, such that $\hat{U}_n |\Psi_n\rangle = |n\rangle$. Therefore, the state preparation scheme $\hat{U}_n |\Psi_n\rangle = |n\rangle$ results to a lower MMSE for all pure states with the same input energy.

We now derive the expression of the MMSE for a Fock state input and the optimal measurement that attains it. In general, the main difficulty with the MMSE formulation is to compute analytically the term $\exp(-z\hat{\Gamma}_0)$ required by Eq. \eqref{eq:B}. Since for a Fock input state, the output remains Fock diagonal, per Eq. \eqref{eq:Gammak} $\hat{\Gamma}_0$ is Fock-diagonal, rendering $\exp(-z\hat{\Gamma}_0)$ Fock-diagonal as well. In Appendix \ref{app:MMSE} we derive the MMSE,
\begin{eqnarray}
 \nonumber   \delta^{|n\rangle}&=&\delta^{|n\rangle}_{\text{LB}}\\
  \label{eq:FockMMSE}  &=& q \tau_0^2+(1-q)\tau_1^2-\sigma^{(n)}(q,\tau_0,\tau_1),
\end{eqnarray}
where,
\begin{eqnarray}
 \nonumber \sigma^{(n)}(q,\tau_0,\tau_1) = \sum_{l=0}^n \frac{[q \tau_0 e_l^{(n)}(\tau_0)+(1-q) \tau_1 e_l^{(n)}(\tau_1)]^2}{q e_l^{(n)}(\tau_0)+(1-q) e_l^{(n)}(\tau_1)}
\end{eqnarray}
and
\begin{eqnarray}
 \label{eq:eigvalue}   e_l^{(n)}(\tau)= \binom{n}{l} \tau^{n-l} (1-\tau)^l.
\end{eqnarray}

Since the $\hat{\Gamma}_k$ operators are diagonal on the Fock basis, the operator $\hat{B}$ is also Fock-diagonal, therefore the optimal projective measurement consists of projections on the Fock basis, i.e., PNR detection. In Appendix \ref{app:B}, we find the form of the $\hat{B}$ operator,
\begin{eqnarray}
 \label{eq:BTPP}   \hat{B}=\sum_{l=0}^n b_l^{(n)}(q,\tau_0,\tau_1) |n-l\rangle \langle n-l|,
\end{eqnarray}
where,
\begin{eqnarray}
 \nonumber  b_l^{(n)}(q,\tau_0,\tau_1)=\frac{q \tau_0 e_l^{(n)}(\tau_0)+(1-q) \tau_1 e_l^{(n)}(\tau_1)}{q e_l^{(n)}(\tau_0)+(1-q) e_l^{(n)}(\tau_1)}.
\end{eqnarray}
We note that even though the symmetric logarithm derivative found in \cite{Adesso2009} (derived within the Fisherian context) and the $\hat{B}$ presented here are both Fock-diagonal, their exact form is different, i.e., their eigenvalues are different, reflecting the incomparability of the Bayesian and the Fisherian approaches. Moreover, as we discuss in Section \ref{sec:concl}, it is not obvious what is the optimal state for arbitrary prior PDFs.

\section{Non-integer mean photon number}\label{sec:non-int}
We now consider the input mean photon number $\bar{n}$ as a real non-negative, not necessarily an integer. For this case, we provide numerical evidence that the optimal state has the form,
\begin{eqnarray}
\label{eq:in-between}
|\Phi_{\bar{n}}\rangle =|a(\bar{n})||\lceil \bar{n} \rceil-1\rangle + |c(\bar{n})||\lceil \bar{n} \rceil\rangle
\end{eqnarray}
where,
\begin{eqnarray}
|c(\bar{n})|&=&\sqrt{1-\lceil \bar{n} \rceil+\bar{n}}\\
|a(\bar{n})|&=&\sqrt{1-|c(\bar{n})|^2}
\end{eqnarray}
and $\lceil \bar{n} \rceil$ is the ceiling of $\bar{n}$. We refer to the state of Eq. \eqref{eq:in-between} as \emph{in-between state} as it is a superposition of the two nearest Fock states for a given $\bar{n}$ and it reverts to a Fock state when $\bar{n}$ is an integer.

For our numerical simulations, we create random states that satisfy the following conditions,
\begin{eqnarray}
\label{eq:num_state}
|\Psi_{\bar{n}}\rangle &=&\sum_{n=0}^{N} d_n|n\rangle,
\\ \sum_{n=0}^{N}|d_n|^2 &=& 1,
\\ \sum_{n=0}^{N}n |d_n|^2 &=& \bar{n},
\end{eqnarray}
for $d_n\in \mathbb{C}$ and by choosing $N$ and $\bar{n}$. For the two-point prior PDF \eqref{eq:TwoPointPrior}, we verify that for integer $\bar{n}$ the optimal state is the Fock state with the same photon number. For real $\bar{n}$, our numerical results support the conjecture that the optimal state has the form of the state \eqref{eq:in-between}, up to a phase, i.e.,
\begin{eqnarray}
\label{eq:GlobalInBetween1} |\Phi'_{\bar{n}}\rangle = e^{i \phi \hat{n}}|\Phi_{\bar{n}}\rangle.
\end{eqnarray}
By removing the global phase, the state in Eq. \eqref{eq:GlobalInBetween1} can be rewritten as,
\begin{eqnarray}
\label{eq:GlobalInBetween2} |\Phi'_{\bar{n}}\rangle = |a(\bar{n})||\lceil \bar{n}  \rceil-1\rangle + |c(\bar{n})| e^{i \phi} |\lceil \bar{n} \rceil\rangle,
\end{eqnarray}
which in turn is (up to a global phase and by setting $\phi=\theta_2-\theta_1$) equal to the general state $\sqrt{1-|c(\bar{n})|^2} e^{i \theta_1}|\lceil \bar{n}  \rceil-1\rangle + |c(\bar{n})| e^{i \theta_2} |\lceil \bar{n} \rceil\rangle$. However, the phase operator in Eq. \eqref{eq:GlobalInBetween1} commutes with the pure-loss channel \cite{Holevo2012}. Therefore, the state of Eq. \eqref{eq:in-between} interacts with the pure-loss, i.e., it picks up the unknown variable $\tau$, and then the phase operator is applied on the channel's output. Since the MMSE is invariant under unitary operators that do not depend on the unknown variable and were applied after the unknown variable has been inscribed on the the state \footnote{This is proven by using Eqs. \eqref{eq:Gammak} and \eqref{eq:B} and showing that the transformed $\hat{\Gamma}_k$ operators are $e^{i \phi \hat{n}}\hat{\Gamma}_k e^{-i \phi \hat{n}}$ and the transformed $\hat{B}$ operator is $e^{i \phi \hat{n}}\hat{B} e^{-i \phi \hat{n}}$. Then, using Eq. \eqref{eq:delta1} and the cyclic permutation property of operators within a trace one can show that the unitary will have no effect on the MMSE}, the MMSE for the states of Eqs. \eqref{eq:in-between} and \eqref{eq:GlobalInBetween2} is the same.

\begin{figure}
\centering
\includegraphics[width=0.45\textwidth]{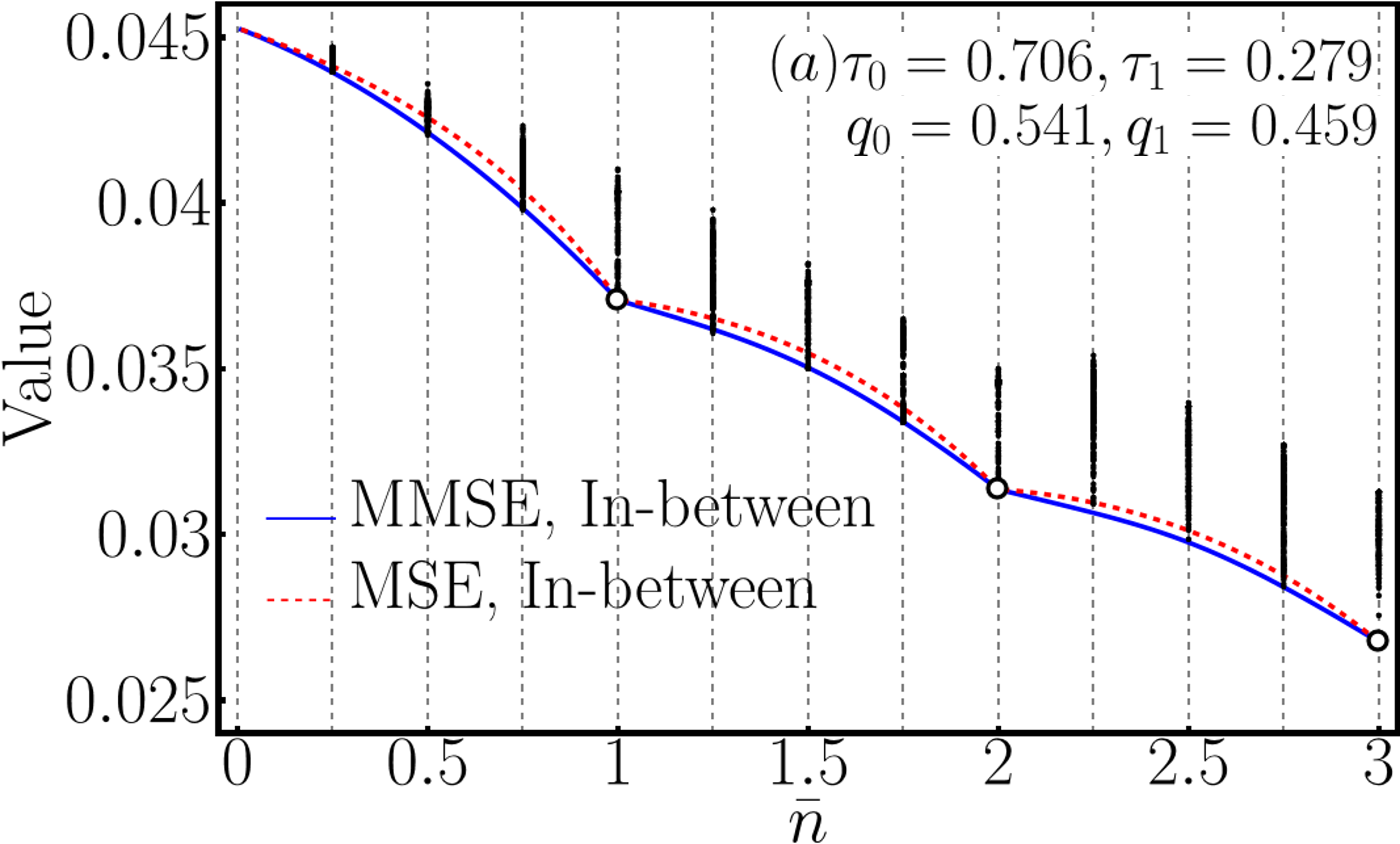}
\includegraphics[width=0.45\textwidth]{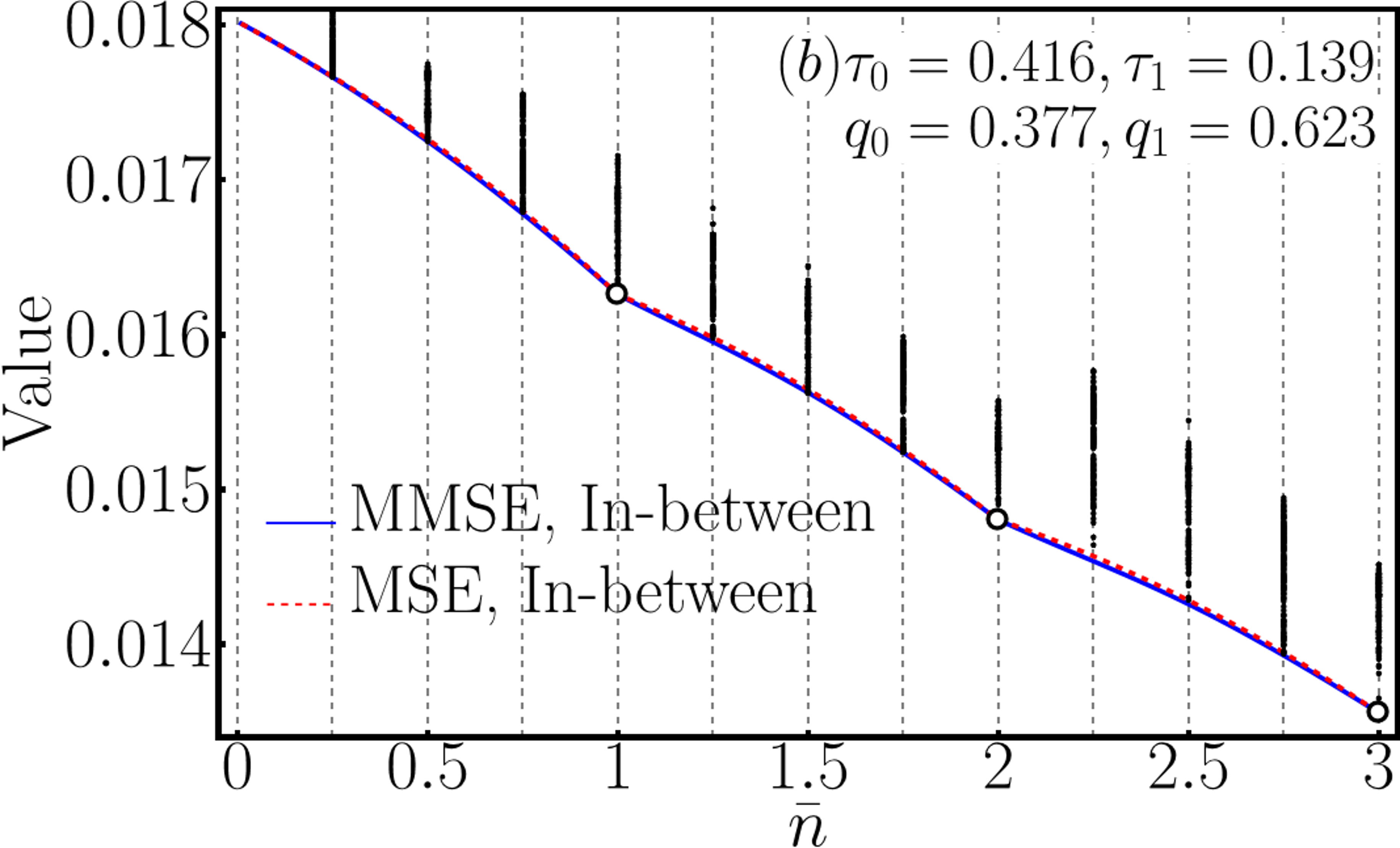}

\caption{For the two-point prior PDF, we plot the MSE of states with mean photon number $\bar{n}$. The blue lines represent the in-between states of Eq. \eqref{eq:in-between}, the red dashed lines represent the MSE by choosing PNR detection, the black solid dots represent random states with fixed $\bar{n}$, and the black empty dots represent Fock states.
(a) Transmissivities $\tau_0=0.706$, $\tau_1=0.279$, and prior PDF with parameters $q_0=0.541$, $q_1=0.459$; (b) transmissivities $\tau_0=0.416$, $\tau_1=0.139$, and prior PDF with parameters $q_0=0.377$, $q_1=0.623$.}
\label{fig_two_point_MMSE}
\end{figure}

We note here that we performed our numerical computations by letting the Fock coefficients of our random states to be complex numbers (see Eqs. \eqref{eq:num_state}). In Fig. \ref{fig_two_point_MMSE} we show an example of our numerical results for $N=4$ (i.e. a superposition of five Fock states) and $200$ states for each fixed $\bar{n}$. The blue lines represent the MSE of the states of Eq. \eqref{eq:in-between}, the black solid dots represent random pure states with fixed mean photon number $\bar{n}$, and the black empty dots represent Fock states. We see that the black dots are above the blue lines.

\subsection{Sub-optimal measurement}
For the in-between states of Eq. \eqref{eq:in-between}, $\hat{B}$ is not diagonal and we were not able to find a useful (in the sense of feasibility) optimal measurement via analytical and numerical means. However, we examine the behavior of PNR detection: We find that, albeit sub-optimal (except if $\bar{n}\in \mathbb{N}$, where it becomes optimal), its values are close to the conjectured MMSE (i.e. the MSE for the sates of \eqref{eq:in-between}). Also, we see that the MSE for PNR detection comes closer to the conjectured MMSE as the differences $|\tau_0-\tau_1|$ and 
$|q_0-q_1|$ become smaller; as said differences go to zero, the two-point prior PDF approaches a delta function, for which all MSEs are zero. We show our results in Fig. \ref{fig_two_point_MMSE}. For PNR detection, the MSE $\Tilde{\delta}$ can be computed as follows,
\begin{eqnarray}
\label{eq:MSE}
\Tilde{\delta}=\int_0^1 d\tau \sum_{k=0}^{n} P(\tau)P(k|\tau)\left( \Pi_k-\tau \right)^2,
\end{eqnarray}
where $k$ is the detected photon number, $P(\tau)$ is the prior probability, $P(k|\tau)$ is the conditional probability, and $\Pi_k$ is the expected value,
\begin{eqnarray}
\nonumber \Pi_k&=&\int_0^1 d\tau P(\tau|k) \tau
\label{eq:MSE_pik1}\\&=& \frac{\int_0^1 d\tau  P(k|\tau)P(\tau) \tau}{\int_0^1 d\tau P(k|\tau)P(\tau)},
\end{eqnarray}
where we used Bayes' rule for $P(\tau|k)$.
For the two-point prior distribution of Eq. \eqref{eq:TwoPointPrior} and the output state $\hat{\rho}_{\bar{n}}(\tau)$ if the input is the state of Eq. \eqref{eq:in-between}, we find,
\begin{eqnarray}
\Pi_k=\frac{q_0\tau_0P(k|\tau_0)+(1-q_0)\tau_1P(k|\tau_1)}{q_0P(k|\tau_0)+(1-q_0)P(k|\tau_1)},
\end{eqnarray}
\begin{eqnarray}
P(k|\tau)=\langle k|\hat{\rho}_{\bar{n}}(\tau)|k\rangle=P^{(k,\tau)}_1+P^{(k,\tau)}_2
\end{eqnarray}
where,
\begin{eqnarray}
 \nonumber   P^{(k,\tau)}_1&=&|c(\bar{n})|^2 (1-\tau)^{\lceil \bar{n} \rceil-k}\tau^k \binom{\lceil \bar{n} \rceil}{\lceil \bar{n} \rceil-k},\\
 \nonumber   P^{(k,\tau)}_2&=&(1-|c(\bar{n})|^2)(1-\tau)^{n-k-1}\tau^k \binom{\lceil \bar{n} \rceil-1}{k},
\end{eqnarray}
and $0\leq k \leq \lceil \bar{n} \rceil$ with the convention $\binom{\lceil \bar{n} \rceil-1}{\lceil \bar{n} \rceil}=0$.

\section{beta distribution as prior PDF}\label{sec:beta}

\begin{figure}
\centering
\includegraphics[width=0.45\textwidth]{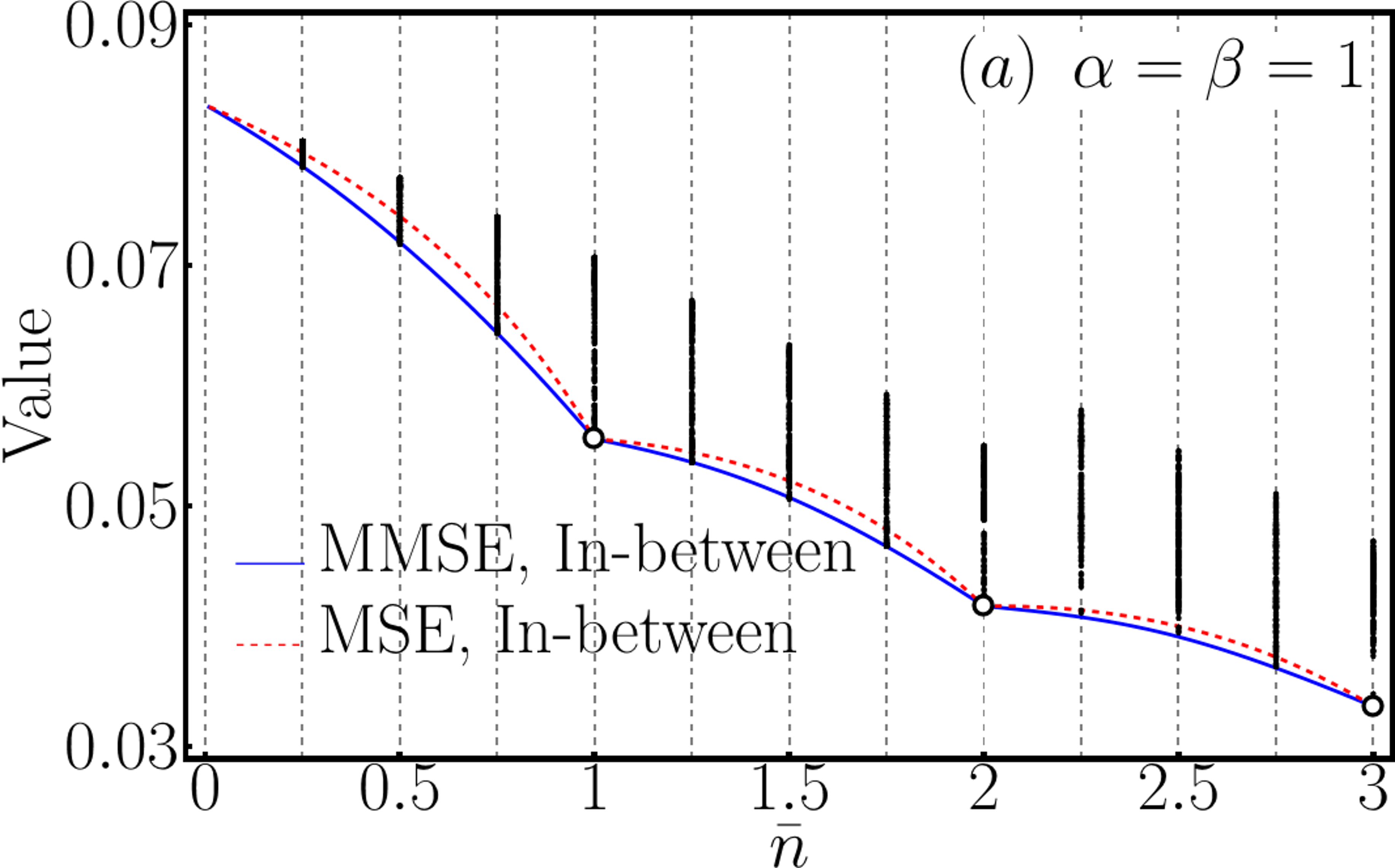}
\includegraphics[width=0.45\textwidth]{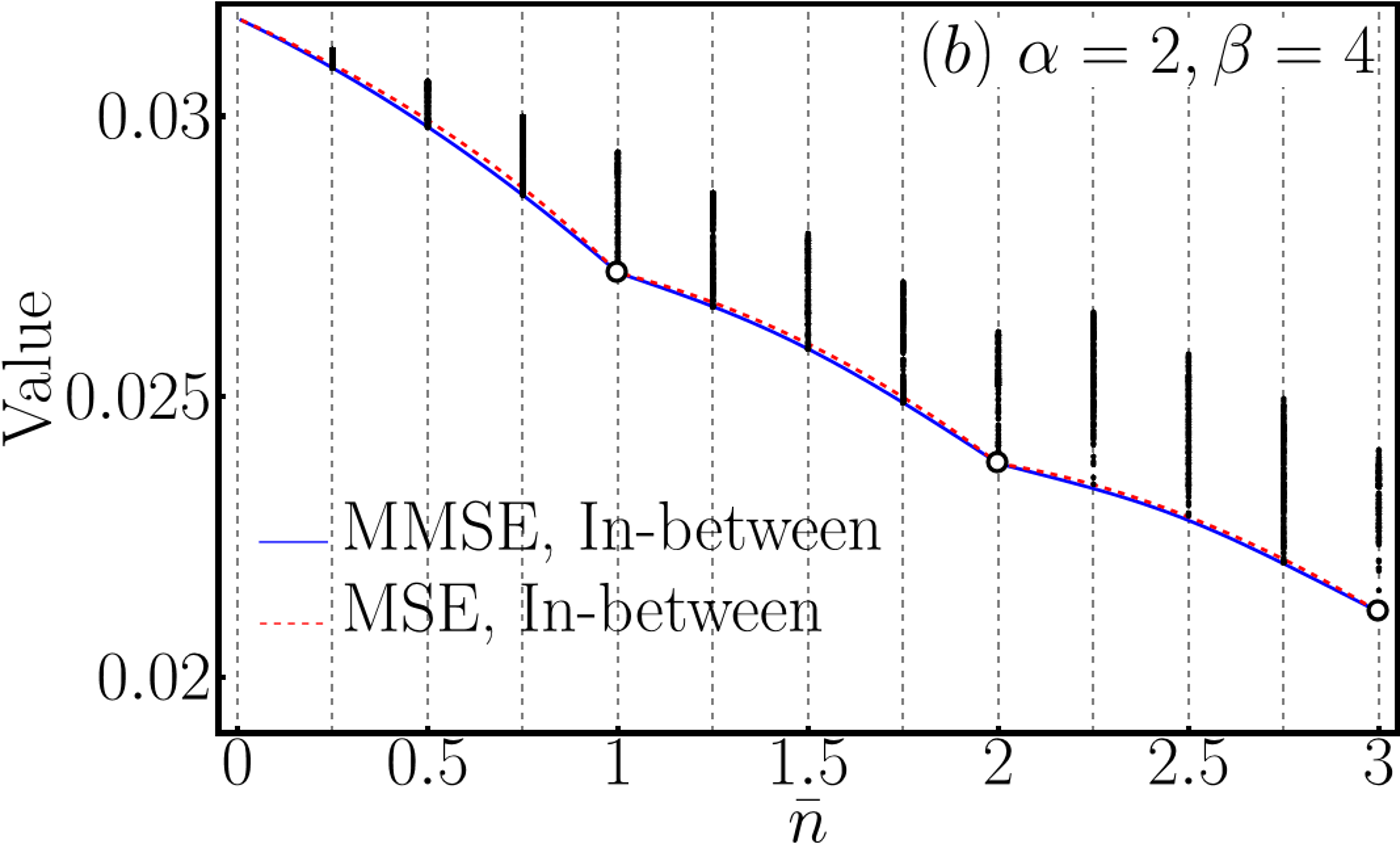}

\caption{For the beta prior PDF, we plot the MSE as a function of $\bar{n}$. The blue lines represent the MSE of in-between states of Eq. \eqref{eq:in-between}, the red dashed lines represent the MSE for PNR detection, the black solid dots represent 200 random states for each fixed $\bar{n}$; the black empty dots represent Fock states. For all random states we used random superpositions of the first five Fock states.
(a) parameters $\alpha=\beta=1$ (uniform distribution); (b) parameters $\alpha=2$, $\beta=4$.}
\label{fig_beta_MMSE}
\end{figure}

We now consider the prior PDF of  Eq.\eqref{eq:betaPrior}. For that case, we can analytically calculate the MSE for a Fock state as input to the pure-loss channel (see Appendix \ref{app:MMSE}),
\begin{eqnarray}
\label{eq:beta_Fock}
\delta=\frac{\alpha \beta}{(\alpha+\beta) (\alpha+\beta+1) (\alpha+\beta+\bar{n})}.
\end{eqnarray}
For non-integer mean photon number and non-Fock states, we compute the MSE numerically using the same methods as in Section \ref{sec:non-int}. The numerical results indicate that the optimal input states with non-integer mean photon number, have the form of Eq. \eqref{eq:in-between}, which means that for $\bar{n} \in \mathbb{N}$ the optimal states are Fock states. We show our results in Fig. \ref{fig_beta_MMSE}.

\section{Discussion and Conclusions}\label{sec:concl}
For the task of sensing the transmissivity of a pure-loss channel, the QFI for a Fock state $|n\rangle$ has been found to be \cite{Adesso2009}\footnote{Taking into account the different parameterization between this work and \cite{Adesso2009}, the results match.},
\begin{eqnarray}
    \label{eq:QFItau} F(\tau) = \frac{n}{\tau (1-\tau)},
\end{eqnarray}
also $F(\tau)$ is attainable by photon-counting. Let us discuss and revisit the main results of present work.

\begin{figure}[t]
\centering
\includegraphics[width=0.45\textwidth]{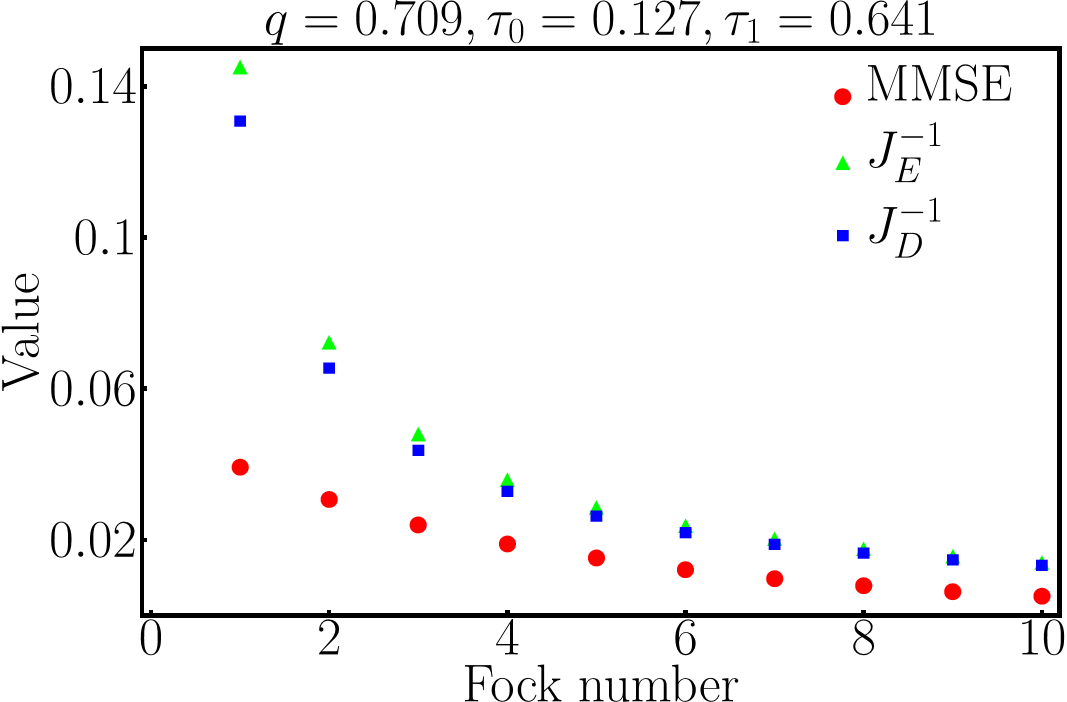}
\caption{For Fock states and the two-point prior PDF with parameters $q=0.79,\ \tau=0.127,\ \tau=0.641$, we plot the Fisherian-inspired Bayesian bounds (triangles and squares) and the MMSE. The MMSE is the lowest one and by definition attainable.}
\label{fig:FITwoPoint}
\end{figure}

\emph{Two-point prior PDF:} For the prior PDF of Eq. \eqref{eq:TwoPointPrior}, the Fisherian-inspired, Bayesian bound of Eq. \eqref{eq:JEinv} gives,
\begin{eqnarray}
J_E^{-1} = \frac{\tau_1 (1-\tau_1)+q (\tau_0-\tau_1)(1-\tau_0-\tau_1)}{n}
\end{eqnarray}
which in general is non-negative, including the case where $q=0$ (or $q=1$) for which the prior PDF becomes a delta function and therefore the Personick MMSE gives zero. Therefore, even if it is known that the Fock state is optimal in the Fisherian approach, the Fisherian-inspired bounds have room for improvement begging the question of what the optimal states and measurements are, that we considered in this work. Surprisingly, we found that the optimal state is unique and is the Fock state, while the optimal measurement is photon-counting. Moreover, we conjectured the form of the optimal states for $\bar{n}>0$ and we examined the performance of photon-counting for such states. For completeness, we also give the formulas,
\begin{eqnarray}
    J_D = n \left[\frac{q}{\tau_0 (1-\tau_0)}+\frac{1-q}{\tau_1(1-\tau_1)}\right]
\end{eqnarray}
for a Fock state $|n\rangle$ as input, while for $J_P$ the integral is problematic since it involves derivative and inverse on the delta function. However, we believe that $J_P$ diverges to $+\infty$, meaning that $J_B \rightarrow +\infty$, which results to a in general non-attainable bound. We present an example of our computations in Fig. \ref{fig:FITwoPoint}, where instead of $J_B^{-1}$, we plot $J_D^{-1}$ which is higher than the Personick MMSE.

\begin{figure}[t]
\centering
\includegraphics[width=0.45\textwidth]{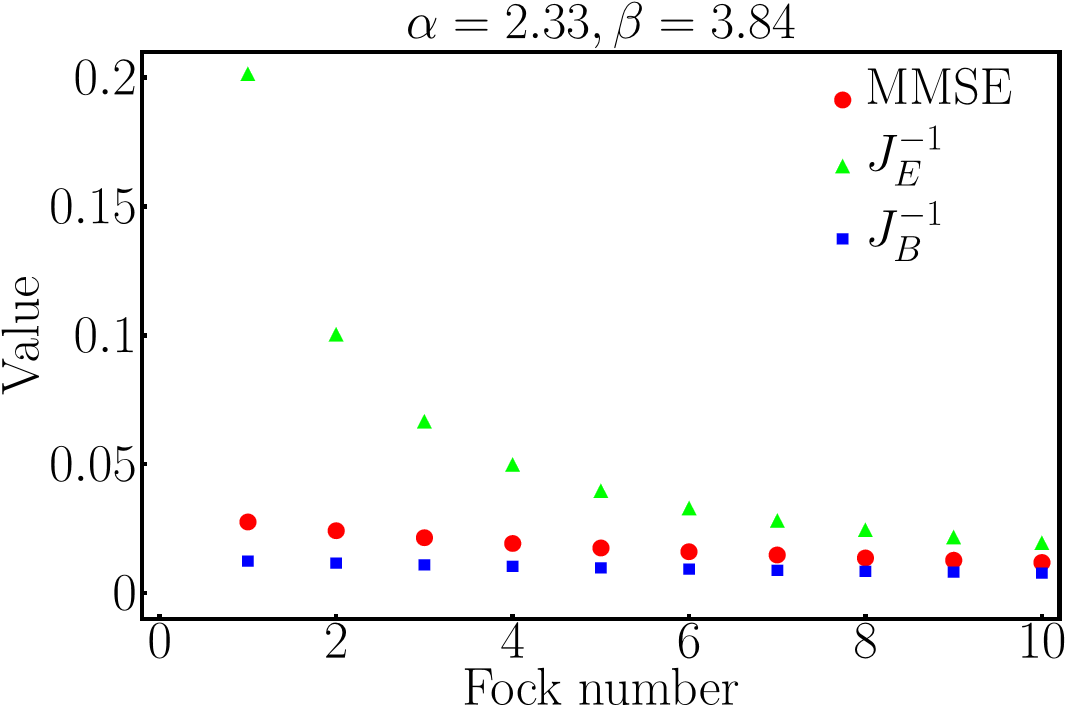}
\caption{For Fock states and the beta prior PDF with parameters $\alpha=2.33,\ \beta=3.84$, we plot the Fisherian-inspired Bayesian bounds (triangles and squares) and the MMSE. The MMSE is the lowest attainable quantity, emphasizing the fact that $J_B^{-1}$ is not attainable by a measurement.}
\label{fig:FIbeta}
\end{figure}

\emph{Beta prior PDF:} For that case we find,
\begin{eqnarray}
    J_E^{-1} &=& \frac{\alpha \beta}{n (\alpha+\beta)(\alpha+\beta+1)}\\
  \label{eq:JBbeta}  J_B &=& \mathcal{C}(\alpha,\beta)\left[n f(\alpha,\beta)+g(\alpha,\beta)) \right]
\end{eqnarray}
where $\mathcal{C}(\alpha,\beta)=\mathcal{B}(\alpha-2,\beta-2)/\mathcal{B}(\alpha,\beta)$, $f(\alpha,\beta)=(\alpha-2)(\beta-2)$, and $g(\alpha,\beta)=(\alpha-1)(\beta-1)(\alpha+\beta-4)$. The computation of $J_P$ required for Eq. \eqref{eq:JBbeta} converges only if $\alpha>2$ and $\beta>2$, reflecting the problematic nature of the Fisherian-inspired bounds. We present an example of our computations in Fig. \ref{fig:FIbeta}.

We note that for the beta prior PDF, we were not able to prove analytically that the Fock states or the in-between states of Eq.~\eqref{eq:in-between} states are optimal, even though our numerical computations support such conjecture. To prove this, following the lower bound approach of Eq.~\eqref{eq:deltaLB} and for $\bar{n}\in \mathbb{N}$, one should show that the lower bound is a function of the mean input photon number only; That would mean that even if states other than Fock, albeit with the same mean-photon number, satisfy Eq. \eqref{eq:comm1}, these states give the same MMSE. Another route would be to prove that for all prior PDFs, Eq. \eqref{eq:comm1} necessarily gives $[\hat{\rho}(\tau),\hat{\rho}(\tau')]=0$, however we note that this seems to be an onerous task (and probably non-true). 

Aforesaid difficulties make the genuine Bayesian approach interesting and non-trivial. Future directions, could include the study of optimal Gaussian states and optimal measurements for sensing the transmissivity of a pure-loss channel. For that case, the difficulty in calculating the MMSE is found in the fact that even if $\hat{\rho}(\tau)$ is Gaussian, $\hat{\Gamma}_0$ is not. Other future directions are sensing the transmissivity including a thermal environment, and sensing multiple parameters of a general Gaussian transformation (or channel) by devising algorithms in the same spirit as in \cite{Lee2022}.

\bibliography{bibliogr.bib}

\newpage
\appendix
\section{Fock-diagonal outputs of pure-loss channel necessarily correspond to Fock-diagonal inputs}\label{app:FockDiagonal}
\renewcommand{\thesubsection}{\arabic{subsection}}
\def\theequation{A\arabic{equation}}
\setcounter{equation}{0}
\renewcommand{\thefigure}{A\arabic{figure}}    
\setcounter{figure}{0}
Although the result of this appendix may be known, we could not find a formal proof.  Thus, we provide one here.

We consider a generic single-mode density operator written on the Fock basis,
\begin{eqnarray}
    \label{eqapp:rho} \hat{\rho}(t)=\sum_{n,m} c_{n,m}(t) |n\rangle \langle m|.
\end{eqnarray}
Then, Eq. \eqref{eq:MasterEq} can be written as,
\begin{eqnarray}
\nonumber \dot{c}_{m,n} &=& \frac{\gamma}{2}[2 \sqrt{(m+1)(n+1)} c_{m+1,n+1}\\
\label{eqapp:MasterEq} &&-(m+n)c_{m,n}],
\end{eqnarray}
where $c_{m,n}(t)\equiv c_{m,n}$ are time-dependent (unless otherwise specified) and the dot represents derivative with respect to time.

We can rewrite Eq. \eqref{eqapp:MasterEq} as $\dot{\mathbf{c}}=\mathbf{A}\mathbf{c}$, where $\mathbf{c}$ is a vectorized version of the matrix with elements $c_{m,n}$, and $\mathbf{A}$ is a matrix which by inspection of Eq. \eqref{eqapp:MasterEq} is upper triangular.
For that reason, we choose to rewrite $c_{m,n}$ as $c_{n+l,n}$, i.e.,
\begin{eqnarray}
  \nonumber  \dot{c}_{n+l,n}&=&\gamma \sqrt{(n+1)(n+l+1)} c_{n+l+1,n+1}\\
 \label{eqapp:MasterEq2}  && -\gamma\left(n+\frac{l}{2}\right) c_{n+l,n},
\end{eqnarray}
where $l=0,1,2,\ldots$, and $l=0$ gives the time evolution of the diagonal elements of Eq. \eqref{eqapp:rho}. For any fixed $l$, we denote a vector,
\begin{eqnarray}
  \label{eqapp:clVec}  \mathbf{c}^{(l)}=\begin{pmatrix}
        c_{0+l,0}\\c_{1+l,1}\\ \vdots \\ c_{n+l,n} \\ \vdots
    \end{pmatrix}.
\end{eqnarray}
Under these considerations, we write Eq. \eqref{eqapp:MasterEq2},
\begin{eqnarray}
 \label{eqapp:EvolOfclVec}   \begin{pmatrix}
        \dot{\mathbf{c}}^{(l=0)}\\ \dot{\mathbf{c}}^{(l=1)}\\ \dot{\mathbf{c}}^{(l=2)}\\ \vdots
    \end{pmatrix} =\mathbf{A} \begin{pmatrix}
        \mathbf{c}^{(l=0)}\\ \mathbf{c}^{(l=1)}\\ \mathbf{c}^{(l=2)}\\ \vdots
    \end{pmatrix},
\end{eqnarray}
where $\mathbf{A}$ is a block-diagonal matrix or the form,
\begin{eqnarray}
  \label{eqapp:Amatrix}  \mathbf{A} = \begin{pmatrix}
        \mathbf{A}^{(l=0)} & & & \\
        & \mathbf{A}^{(l=1)} & & \\
        & & \mathbf{A}^{(l=2)} &  \\
         & & &  \ddots 
    \end{pmatrix},
\end{eqnarray}
while all empty entries are equal to zero. The matrix $\mathbf{A}^{(l)}$ has the form,
\begin{eqnarray}
 \label{eqapp:dplusud}   \mathbf{A}^{(l)}=\mathbf{A}_{\text{D}}^{(l)}+\mathbf{A}_\text{UD}^{(l)},
\end{eqnarray}
where the matrix $\mathbf{A}_\text{D}^{(l)}$ is diagonal,
\begin{eqnarray}
 \mathbf{A}_{\text{D}}^{(l)} = -\gamma \mathbf{D},
\end{eqnarray}
where $\mathbf{D}=\text{diag}\left[  \frac{l}{2}+0, \frac{l}{2}+1, \frac{l}{2}+2,\ldots\right]$,
and the matrix $\mathbf{A}_\text{UD}^{(l)}$ is upper-diagonal (i.e. the only non-zero elements are in its upper diagonal),
\begin{eqnarray}
 \mathbf{A}_{\text{UD}}^{(l)} = \gamma \mathbf{D}_u ,
\end{eqnarray}
where $\mathbf{D}_u = \text{udiag}\left[ \sqrt{l+1}, \sqrt{2(l+2)}, \sqrt{3(l+3)},\ldots\right],$ where by $\text{udiag}$ we denote an upper diagonal matrix.

The solution to each differential equation (i.e. for each $l$) represented in Eq. \eqref{eqapp:EvolOfclVec} is given by,
\begin{eqnarray}
    \label{eqapp:Solution} \mathbf{c}^{(l)} = \exp\left(\mathbf{A}^{(l)} t\right) \mathbf{c}^{(l)}_0,
\end{eqnarray}
where $\mathbf{c}^{(l)}$ is time-dependent and $\mathbf{c}^{(l)}_0\equiv\mathbf{c}^{(l)}(t=0)$ is given by the initial condition of Eq. \eqref{eq:InitialCond} on the Fock basis. 

We note that since $\mathbf{A}^{(l)}$ is upper-triangular, the matrix $\exp\left(\mathbf{A}^{(l)} t \right)$ is upper triangular as well. Let us work out the specific form of $\exp\left(\mathbf{A}^{(l)} t \right)$. We have,
\begin{eqnarray}
    \exp\left(\mathbf{A}^{(l)} t \right)=\exp\left(\mathbf{A}_{\text{D}}^{(l)} t +\mathbf{A}_{\text{UD}}^{(l)} t\right).
\end{eqnarray}
To apply the Zassenhaus formula \cite{Magnus1954}, we first calculate the commutator,
\begin{eqnarray}
    \mathbf{C}_N = [\mathbf{A}^{(l)}_{\text{D}},\mathbf{C}_{N-1}],
\end{eqnarray}
with $\mathbf{C}_0=\mathbf{A}^{(l)}_{\text{UD}}$. We find,
\begin{eqnarray}
    \mathbf{C}_N = \gamma^N \mathbf{A}^{(l)}_{\text{UD}}.
\end{eqnarray}
We also find,
\begin{eqnarray}
    [\mathbf{A}^{(l)}_{\text{UD}},\mathbf{C}_N]=0.
\end{eqnarray}
Applying the Zassenhaus formula we get,
\begin{eqnarray}
 \nonumber   \exp\left(\mathbf{A}^{(l)} t \right)&=&\exp\left(\mathbf{A}_{\text{D}}^{(l)} t\right)\\
 \nonumber &&\times\exp\left(\sum_{j=1}^\infty \frac{t^j}{j!}(-1)^{j+1} \gamma^{j-1} \mathbf{A}_{\text{UD}}^{(l)}\right),
\end{eqnarray}
which gives,
\begin{eqnarray}
     \label{eqapp:ExpAt}   \exp\left(\mathbf{A}^{(l)} t \right)=\exp\left(\mathbf{A}_{\text{D}}^{(l)} t\right)\exp\left(\mathbf{S}^{(l)}\right),
\end{eqnarray}
where,
\begin{eqnarray}
 \nonumber  \mathbf{S}^{(l)} = (1-e^{-t \gamma})
 \mathbf{D}_u,
\end{eqnarray}
from which we find the $ij$ (we enumerate as $i,j=1,2,\ldots$ everywhere) element of $\exp\left(\mathbf{S}^{(l)}\right)$ for $j\geq i$,
\begin{eqnarray}
 \nonumber   \exp\left(\mathbf{S}^{(l)}\right)_{ij} = \frac{(1-e^{-t \gamma})^{j-i}}{(j-i)!} \prod_{r=i}^{j-1} \sqrt{r (l+r)},
\end{eqnarray}
with the convention $\prod_{r=i}^{i-1} \sqrt{r (l+r)}=1$. For $j < i$ we find $\exp\left(\mathbf{S}^{(l)}\right)_{ij} =0$. Since $\mathbf{A}_{\text{D}}^{(l)}$ is diagonal, the $ij$ element of $\exp\left(\mathbf{A}_{\text{D}}^{(l)}\right)$ is,
\begin{eqnarray}
  \nonumber  \exp\left(\mathbf{A}_{\text{D}}^{(l)}\right)_{ij} = \delta_{ij} \exp\left[-t \gamma \left(\frac{l}{2}+i-1\right)\right]. 
\end{eqnarray}

Multiplying a diagonal matrix with an upper-triangular matrix as per the left hand side of Eq. \eqref{eqapp:ExpAt}, we get an upper-triangular matrix, as expected. Also, all elements of $\exp\left(\mathbf{A}^{(l)} t \right)$ are non-zero for $l \neq 0$ (i.e. for non-diagonal entries of the density operator on the Fock basis), unless when $t \gamma \rightarrow 0$, which is the case of no-interaction ($\tau=e^{-t \gamma}=1$). We demand Eq. \eqref{eqapp:Solution} to be element-wise equal to zero for $l\neq 0$, i.e.,
\begin{eqnarray}
    \label{eqapp:Solution0} \mathbf{c}^{(l\neq 0)} = \exp\left(\mathbf{A}^{(l\neq 0)} t\right) \mathbf{c}^{(l\neq 0)}_0=0.
\end{eqnarray}
In Eq. \eqref{eqapp:Solution0} the upper-triangular matrix $\exp\left(\mathbf{A}^{(l)} t\right)$ multiplies the vector $\mathbf{c}^{(l)}_0$, rendering all coefficients of Eq. \eqref{eqapp:clVec}, starting from the bottom and moving upwards equal to zero. Therefore, if the time-evolved off-diagonal coefficients of Eq. \eqref{eqapp:rho} are zero, then the input density operator is Fock-diagonal. 

If $\tau=e^{-t \gamma}=1$, then the input state is retrieved at the output and trivially such an output is Fock-diagonal if the input is Fock-diagonal.

The converse, i.e., if the input is Fock-diagonal, then the output is also Fock-diagonal is a well known fact referred to as \emph{gauge} or \emph{phase} invariance of the pure-loss channel. 

\section{The optimal projective measurements for Fock states under the two-point prior and beta PDFs}\label{app:B}
\renewcommand{\thesubsection}{\arabic{subsection}}
\def\theequation{B\arabic{equation}}
\setcounter{equation}{0}
\renewcommand{\thefigure}{C\arabic{figure}}    
\setcounter{figure}{0}
For input state $\hat{\rho}_0=|n\rangle \langle n|$, Eq. \eqref{eq:Kraus} gives,
\begin{eqnarray}
 \label{eqapp:rhoDiag}   \hat{\rho}(\tau) = \sum_{l=0}^n e_l^{(n)}(\tau) |n-l\rangle \langle n-l|,
\end{eqnarray}
where $e_l^{(n)}(\tau)$ is given in Eq. \eqref{eq:eigvalue}. For the prior PDF of Eq. \eqref{eq:TwoPointPrior}, from Eq. \eqref{eq:Gammak} we find,
\begin{eqnarray}
 \nonumber   \hat{\Gamma}_k &=& q \sum_{l=0}^n e_l^{(n)}(\tau_0) \tau_0^k |n-l\rangle \langle n-l| \\
 \label{eqapp:GammakTPP}  && + (1-q) \sum_{l=0}^n e_l^{(n)}(\tau_1) \tau_1^k |n-l\rangle \langle n-l|.
\end{eqnarray}
Since $\hat{\Gamma}_0$ is Fock-diagonal we find,
\begin{eqnarray}
 \nonumber   e^{-z \hat{\Gamma}_0} &=& \sum_{l=0}^n \exp\left[-z\left(q e_l^{(n)} (\tau_0) + (1-q) e_l^{(n)}(\tau_1)\right)\right]\\
 \label{eqapp:expGamma0TPP}&& \times |n-l\rangle \langle n-l|
\end{eqnarray}
From Eqs. \eqref{eqapp:GammakTPP} for $k=1$ and \eqref{eqapp:expGamma0TPP},  Eq. \eqref{eq:B} gives Eq. \eqref{eq:BTPP} which is a closed form relation expressed as a finite sum.

The calculation is similar if we use the beta distribution of Eq. \eqref{eq:betaPrior}. In fact, for every density operator of Eq. \eqref{eqapp:rhoDiag} and for \emph{any} prior PDF $P(\tau)$ we can write,
\begin{eqnarray}
 \nonumber   \hat{\Gamma}_k &=& \int_0^1 d\tau P(\tau) \tau^k \sum_{l=0}^n e_l^{(n)}(\tau) |n-l\rangle \langle n-l|\\
 \label{eqapp:GammakGeneric}   &=& \sum_{l=0}^n g_l^{(n,k)}[P] |n-l\rangle \langle n-l|,
\end{eqnarray}
where,
\begin{eqnarray}
 \label{eqapp:smallGamma}   g_l^{(n,k)}[P] = \int_0^1 d\tau P(\tau) \tau^k e_l^{(n)}(\tau).
\end{eqnarray}
The square bracket denotes the functional dependence of $g_l^{(n,k)}[P]$ on $P(\tau)$.

From Eqs. \eqref{eq:B} and \eqref{eqapp:GammakGeneric} for $k=1$, we get,
\begin{eqnarray}
    \hat{B}= \sum_{l=0}^n \frac{g_l^{(n,1)}[P]}{g_l^{(n,0)}[P]} |n-l\rangle \langle n-l|,
\end{eqnarray}
meaning that for a Fock state input, projection on Fock states is always the optimal measurement.\\

\section{The MMSE for Fock states under the two-point prior and beta PDFs}\label{app:MMSE}
\renewcommand{\thesubsection}{\arabic{subsection}}
\def\theequation{C\arabic{equation}}
\setcounter{equation}{0}
\renewcommand{\thefigure}{B\arabic{figure}}    
\setcounter{figure}{0}
Let us consider the two-point prior PDF of Eq. \eqref{eq:TwoPointPrior}. Using Eq. \eqref{eqapp:GammakTPP} for $k=2$ we find,
\begin{eqnarray}
    \label{eqapp:TrGamma2TPP} \text{tr}(\hat{\Gamma}_2) = q \tau_0^2 + (1-q) \tau_1^2.
\end{eqnarray}
From Eqs. \eqref{eq:BTPP} and \eqref{eqapp:GammakTPP} for $k=1$ we find,
\begin{eqnarray}
 \nonumber   \text{tr} (\hat{B}\hat{\Gamma}_1) = \sum_{l=0}^n \frac{[q \tau_0 e_l^{(n)}(\tau_0)+(1-q) \tau_1 e_l^{(n)}(\tau_1)]^2}{q e_l^{(n)}(\tau_0)+(1-q) e_l^{(n)}(\tau_1)},
\end{eqnarray}
which together with Eqs. \eqref{eq:delta1} and \eqref{eqapp:TrGamma2TPP}, gives Eq. \eqref{eq:FockMMSE}.

The calculation is similar if we use the beta distribution of Eq. \eqref{eq:betaPrior} to derive Eq. \eqref{eq:beta_Fock}. In fact, for every density operator of Eq. \eqref{eqapp:rhoDiag} and for \emph{any} prior PDF $P(\tau)$, using Eqs. \eqref{eq:delta1} and \eqref{eqapp:GammakGeneric} for $k=1,2$ we find,
\begin{eqnarray}
 \nonumber   \delta^{|n\rangle}[P] = \sum_{l=0}^n \left[ \left(g_l^{(n,2)}[P]\right)^2-\frac{\left(g_l^{(n,1)}[P]\right)^2}{g_l^{(n,0)}[P]}\right],
\end{eqnarray}
where the square brackets in $\delta^{|n\rangle}[P]$ denote the functional dependence on $P(\tau)$.
\end{document}